**Dielectric spectra of a new relaxor ferroelectric system $Ba_2LnTi_2Nb_3O_{15}$ (Ln=La, Nd)**

S. Kamba, S. Veljko, M. Kempa, M. Savinov, V. Bovtun, P. Vanek, J. Petzelt

Institute of Physics, ASCR, Na Slovance 2, 182 21 Prague 8, Czech Republic

M.C. Stennett, I.M. Reaney, A.R. West

Department of Engineering Materials, The University of Sheffield, Mappin Street, Sheffield UK S1 3JD

**Abstract**

$Ba_2LaTi_2Nb_3O_{15}$ exhibits a smeared maximum of permittivity, characteristic of classic relaxor behaviour, with a peak shift from 185 K at 100 Hz to 300 K at 1 GHz. $Ba_2NdTi_2Nb_3O_{15}$ undergoes a first order ferroelectric phase transition at 389 K and $Ba_2La_{0.5}Nd_{0.5}Ti_2Nb_3O_{15}$ exhibits both a ferroelectric phase transition at 274 K and relaxor behaviour at higher temperatures. All three ceramic systems were investigated across a broad frequency (100 Hz – 1 THz) and temperature (10 – 500 K) range. Slowing down of the dielectric relaxation across the whole investigated temperature range from the THz and microwave regions to below 100 Hz was observed on cooling in the relaxor ferroelectric $Ba_2LaTi_2Nb_3O_{15}$. The mean relaxation frequency obeys the Vogel-Fulcher law with freezing temperature of 40 K. In $Ba_2NdTi_2Nb_3O_{15}$ and $Ba_2La_{0.5}Nd_{0.5}Ti_2Nb_3O_{15}$ the relaxations soften only in the paraelectric phase down to the 1 GHz and 100 MHz range, respectively. Below $T_c$ the relaxations vanish. The origin of the dielectric relaxations is discussed on the basis of structural data.

## Introduction

High and broad maxima in temperature dependence of both components of complex permittivity $\varepsilon^* = \varepsilon' - i\varepsilon''$ and their shift to higher temperatures with raising measuring frequency is a typical feature of relaxor ferroelectrics (RFEs). RFE materials are studied due to both their peculiar physical properties, which are thought to be connected with the presence of polar nanoclusters above and below the temperature of the maximum of permittivity ($T_m$) and their potential technical applications in piezoelectric devices and microelectronics. For a recent overview of RFEs see reviews by Ye[1] and Samara[2, 3].

Most of the RFEs with potential piezoelectric applications are lead-based compounds with the perovskite structure, however there is currently an increased need for more environmental friendly lead-free compounds. Recently a new RFE $Ba_2LaTi_2Nb_3O_{15}$ (BLTN) ceramic with maximum permittivity near 200 K at 10 kHz was reported.[4] The material has the tetragonal tungsten bronze structure, but from the XRD data it was not possible to distinguish between the P4/mbm and P4bm space groups. It should be noted that similar structure have also been seen in other well known RFEs like $Sr_{0.61}Ba_{0.39}Nb_2O_6$ or $Ba_2NaNb_5O_{15}$ (see e.g. Ref.[5])

The aim of this article is to extend the dielectric data of BLTN published in Ref. [4] into the microwave and THz range to better understand the dynamics of the phase transition in this compound. The dielectric spectra of BLTN will be compared with the completely new ferroelectric compounds $Ba_2NdTi_2Nb_3O_{15}$ (BNTN) and $Ba_2La_{0.5}Nd_{0.5}Ti_2Nb_3O_{15}$ (BLNTN).

## Experimental

A new lead-free relaxor ferroelectric system $xBa_2LaTi_2Nb_3O_{15}$ – $(1-x)Ba_2NdTi_2Nb_3O_{15}$, with the tetragonal tungsten bronze structure, has been synthesized. Two end member compositions (x = 0, 1) and one intermediate composition (x = 0.5) have been prepared by solid state ceramic synthesis. The powders were calcined at 1300°C for 8 hours and uniaxially pressed pellets were sintered at 1450°C for 4 hours. Sintered pellet densities were >95% of theoretical. The powders were characterized by X-ray diffraction and the data sets were fully indexed on a tetragonal tungsten bronze unit cell with no evidence of any secondary phase. At room temperature BLTN was assigned the centrosymmetric space group P4/mbm and BNTN the non centrosymmetric space group P4bm.

The low-frequency dielectric response in the 100 Hz – 1 MHz range was obtained using a HP 4192A impedance analyzer with a Leybold He-flow cryostat (operating range 5 – 300 K) and a custom-made furnace (300 – 900 K). Dielectric measurements in the high-frequency (HF) range (1 MHz – 1 GHz) were performed using a computer-controlled HF dielectric spectrometer equipped with a HP 4291B impedance analyzer, a Novocontrol BDS 2100 coaxial sample cell and a Sigma System M18 temperature chamber (operating range 100 – 570 K). The dielectric parameters were calculated taking into account the electromagnetic field distribution in the sample. Time-domain THz transmission spectra were obtained using an amplified femtosecond laser system. Two [110] ZnTe single-crystal plates were used to generate (by optical rectification) and detect (by electro-optic sampling) the THz pulses. The THz technique allows determination of the complex dielectric response $\varepsilon^*(\omega)$ in the range from 3 to 75 $cm^{-1}$ (100 GHz – 2.5 THz); the investigated samples were not transparent enough in this entire range so that the transmission spectra were evaluated in a narrower spectral range only.

Differential scanning calorimetry (DSC) measurements were obtained on Perkin-Elmer Pyris-Diamond DSC calorimeter in the temperature range 100 - 570 K with a temperature rate of 10 K/min.

## Results and discussion

### a) $Ba_2LaTi_2Nb_3O_{15}$ ceramics

The temperature dependence of the complex permittivity in the BLTN ceramic at frequencies between 100 Hz and 400 GHz is shown in Figure 1. Typical relaxor behaviour is seen; the temperature of maximum permittivity shifts from 185 K (at 100 Hz) to 300 K (at 1 GHz). No anomaly was observed in the DSC measurement giving no evidence for a structural phase transition in this material. Dielectric behaviour is qualitatively the same as in Ref.[4], but the value of ε' is more than twice as large. This supports the statement in Ref.[4] that the BLTN ceramics may be non-stoichiometric, and can have variable La to Ba and Ti to Nb ratios and variable O content. This explains the different values of permittivity obtained from different samples. Of more interest is the frequency plot of ε*(ω) below 1 GHz in Figure 2. The wing of the microwave (MW) relaxation is observed at 450 K in the range $10^8$-$10^9$ Hz. The relaxation frequency decreases (see the ε'' increase at 1 GHz) on cooling, reaches 1 GHz at 300 K and further decreases to 1 kHz at 150 K. At lower temperatures the relaxation slows down below the frequency range of measurement. A characteristic feature of all the dielectric spectra is a broadening of the dielectric dispersion on cooling which can be attributed to a broadening of the distribution of relaxation frequencies. This is a consequence of the distribution of energy barriers, for hopping of the disordered ions, that exists in a multi-well potential. XRD data shows that the B sites contain Ti and Nb ions in a statistical way[4] which produce the random fields. The mean relaxation frequency $f_r$ follows the Vogel-Fulcher law

$$f_r = f_0 \exp \frac{-E_a}{k(T - T_{VF})}$$

with the attempt frequency $f_0 = 6.4\ 10^{12}$ Hz, activation energy $E_a = 0.2$ eV and Vogel-Fulcher temperature $T_{VF} = 40$ K (see Figure 3).

Let us now discuss the dielectric dispersion in the THz range, shown in Figure 4. At high temperatures above 525 K only a phonon contribution to ε*(ω) is seen. The increase of permittivity between 625 and 525 K is probably caused by a softening of some of the polar phonons. Below 425 K, the low-frequency part of ε'(ω) starts to increase on cooling due to the high-frequency wing of a MW relaxation, which results also in an increase of the THz dielectric losses. The relaxation could be caused by the dynamics of the polar nanoclusters. This means that the Burns temperature, below which the polar clusters start to appear, lies somewhere between 450 and 525 K. The contribution of the relaxation to the THz dispersion increases on cooling, reaches a maximum at 250 K and than again decreases below 100 K. Above room temperature it is the same relaxation which is seen in Figure 2, but it splits into two components below room temperature. The lower frequency component is responsible for the ε*(T) anomalies at frequencies below 1 GHz and the higher frequency relaxation is responsible for the shifted maximum in ε'(T) and in ε''(T) at 400 GHz (see Figure 1). The lower frequency component probably describes the flipping of the polar clusters, while the higher frequency one expresses the breathing of polar cluster walls.

**b) $Ba_2NdTi_2Nb_3O_{15}$ ceramics**

BNTN ceramics exhibits qualitatively different dielectric behaviour to BLTN. They exhibit a sharp first order ferroelectric phase transition which is shown in Figure 5. Temperature hysteresis was observed between the cooling and heating cycles (heating/cooling rate of 2 K/min) with $T_c$ changing by 26 K in both the dielectric and DSC data. $T_c$ was observed on cooling at 389 K and a change of enthalpy (ΔH) of 2.5 J/g was measured. There was a delay of several days between the low - and the high temperature dielectric measurements, in the range up to 300 kHz therefore the absolute values of ε' and ε'' show a step at 300 K. Nevertheless, the relaxation in ferroelectric phase below 390 K is clearly seen and it is probably connected with domain-wall motion.

A frequency-dependent plot of the complex permittivity in Figure 6 shows the low-frequency wing of the MW relaxation above $T_c$, which gives information about the softening of the relaxation on cooling to $T_c$. Below $T_c$ the strength of this relaxation suddenly decreases, which results in a step in ε'(T). The dielectric loss becomes almost frequency independent. Similar behaviour was observed in other highly disordered systems (polymers, ionic crystals, dipolar glasses etc.) but the most remarkable is seeing this behaviour in relaxor ferroelectrics.[6,7] In the case of BNTN the frequency independent loss is much lower than in the relaxor ferroelectrics. The shape of $\varepsilon^*(\omega)$ can be modelled using a uniform distribution of relaxation frequencies[6] describing the motion of domain walls. The temperature dependence of the dielectric response of BNTN was not measured in the THz range due to the small size of the sample. The room temperature data (see Figure 5), however shows that the phonon contribution to permittivity is about half of the value measured in BLNT ceramics.

### c) $Ba_2La_{0.5}Nd_{0.5}Ti_2Nb_3O_{15}$

BLNTN is solid solution composition containing both $La^{3+}$ and $Nd^{3+}$ previous compositions, therefore it exhibits a mixed behaviour. Relaxor behaviour was observed above room temperature and a first order ferroelectric phase transition at $T_c$ = 274 K (cooling) was observed. DSC revealed temperature hysteresis between the cooling and heating cycles (25 K) and a change of enthalpy of 1.2 J/g at $T_c$. The maximum in the permittivity shifted from 290 K at 100 Hz to 375 K at 1 GHz. MW relaxation was seen above $T_c$, its mean relaxation frequency decreases on cooling to 200 MHz at $T_c$ and than the relaxations vanish on further cooling. Almost frequency independent dielectric losses were observed below 200 K giving evidence about the large disorder in the lattice and a broad distribution of activation energies for motion of disordered ions. Similar behaviour was observed in BNTN ceramics. A striking step down in $\varepsilon^*(T)$ occurs near $T_c$ in the THz frequency range. Corresponding figures will be published separately with the results of far-infrared reflectivity studies of all the ceramics.

## Conclusions

A new ceramic system $Ba_2LnTi_2Nb_3O_{15}$ (Ln=La, Nd), with tetragonal tungsten bronze structure has been synthesized and characterized in the broad spectral range from 100 Hz to 800 GHz. BLTN exhibits ferroelectric relaxor behaviour, BNTN shows a normal ferroelectric phase transition ($T_c$ = 389 K) and BLNTN demonstrates both relaxor and ferroelectric properties ($T_c$ = 274 K). Dielectric relaxation in BLTN softens on cooling from MW frequencies down to less than 100 Hz and obeys the Vogel-Fulcher law with a freezing temperature of 40 K. Pronounced MW relaxations were observed in the paraelectric phases of the BNTN and BLNTN ceramics. Their relaxation frequencies also decrease on cooling to 100 MHz – 1 GHz, but the relaxations vanish from the spectra below $T_c$ and only nearly frequency-independent losses were observed below 200 K.

**Acknowledgments**

The work was supported by the Grant Agency of the Czech Republic (project No. 202/04/0993) and Czech Academy of Sciences (projects Nos. A1010213, AVOZ01-010-914 and K1010104).

**Figure Captions**

Figure 1. Temperature dependence of the real and imaginary part of complex permittivity in BLTN ceramics at various frequencies.

Figure 2. Frequency dependence of the complex permittivity in BLTN ceramics at various temperatures.

Figure 3. Temperature dependence of the mean relaxation frequency in BLTN fitted with the Vogel-Fulcher model.

Figure 4. Complex dielectric response of BLTN ceramics in THz frequency range at selected temperatures. Oscillations in the spectra are noise.

Figure 5. Temperature dependence of the complex permittivity in BNTN ceramics at various frequencies.

Figure 6. Frequency dependence of the complex permittivity in BNTN ceramics at various temperatures.

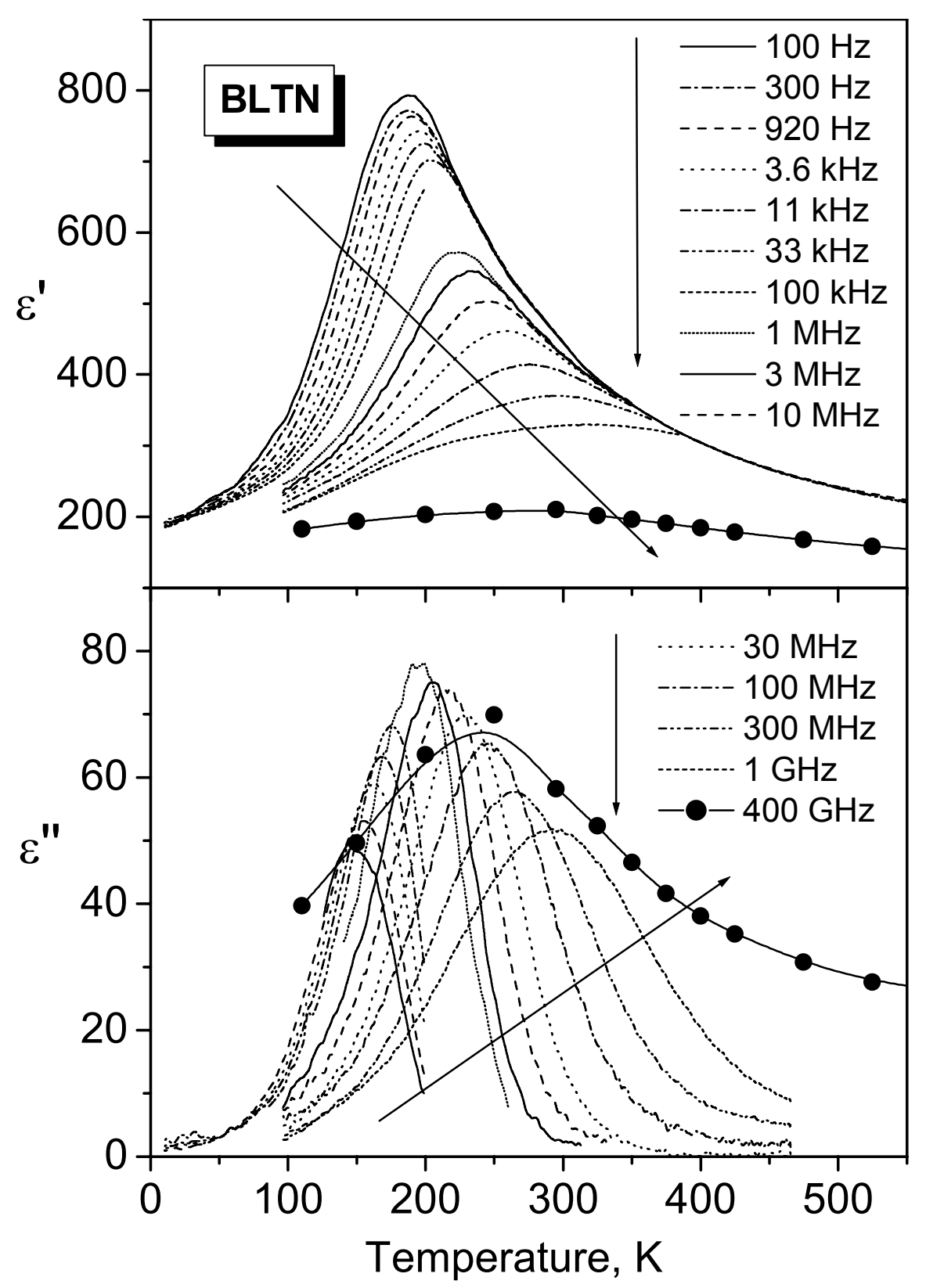

Figure 1

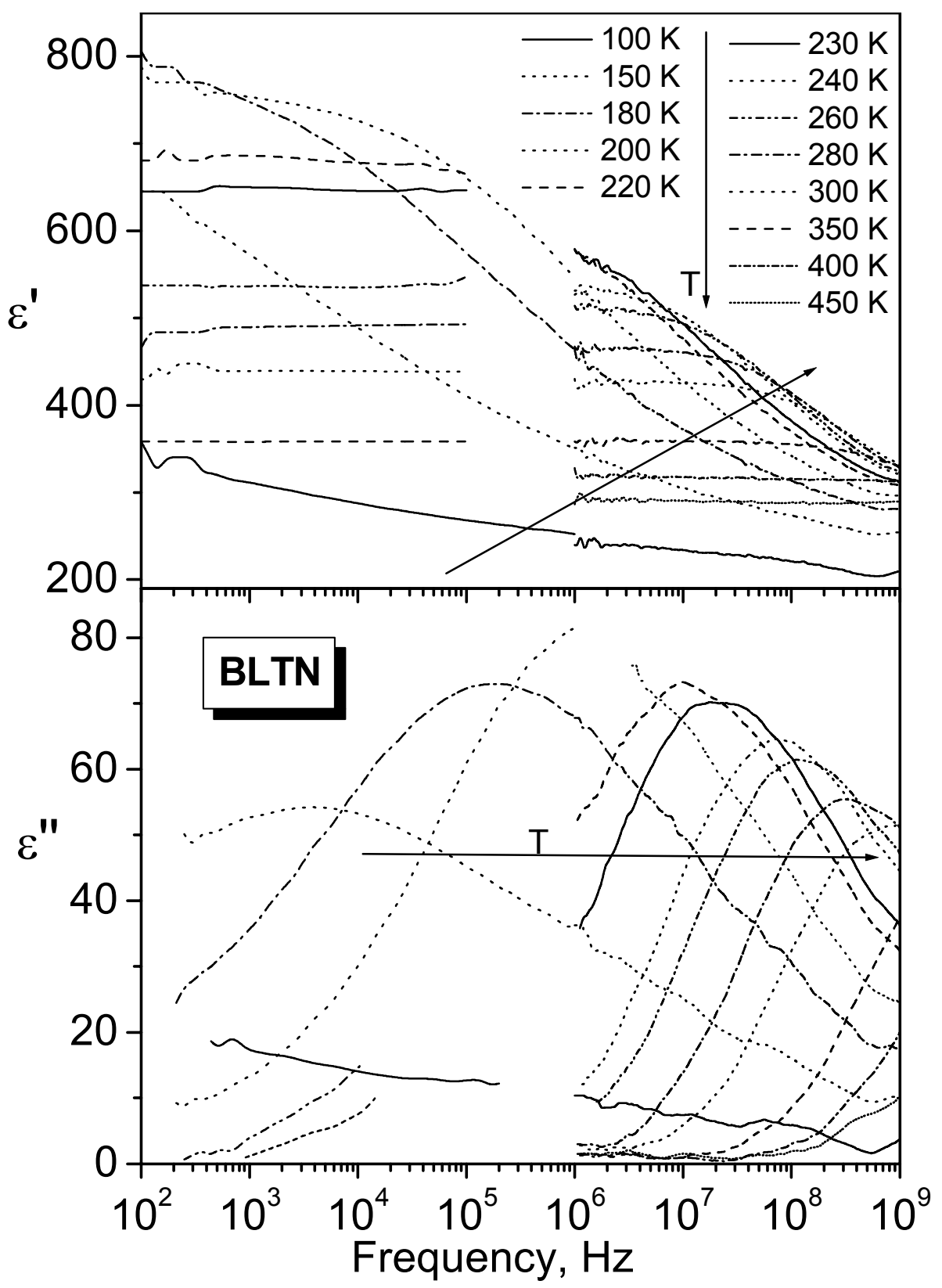

Figure 2

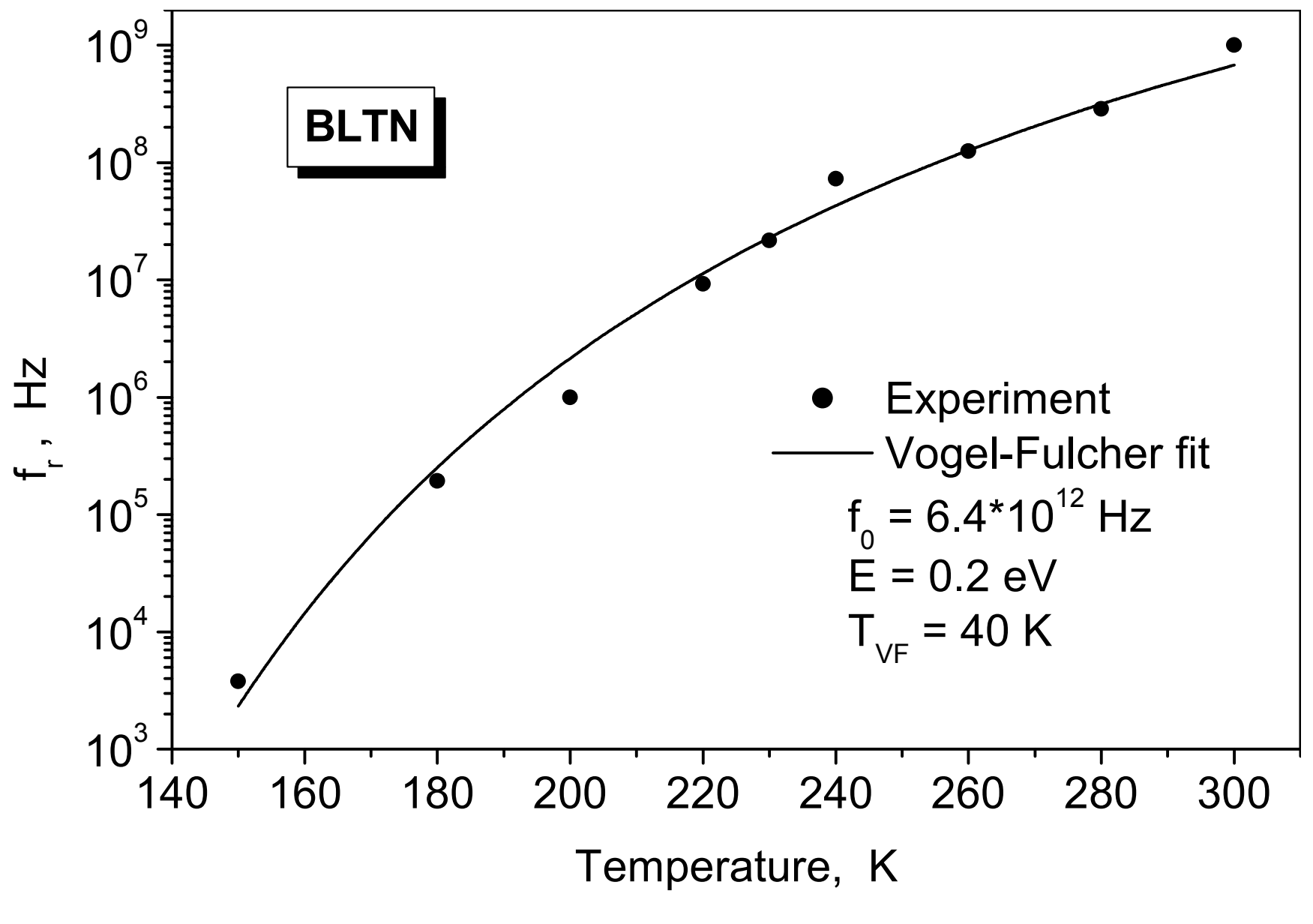

Figure 3

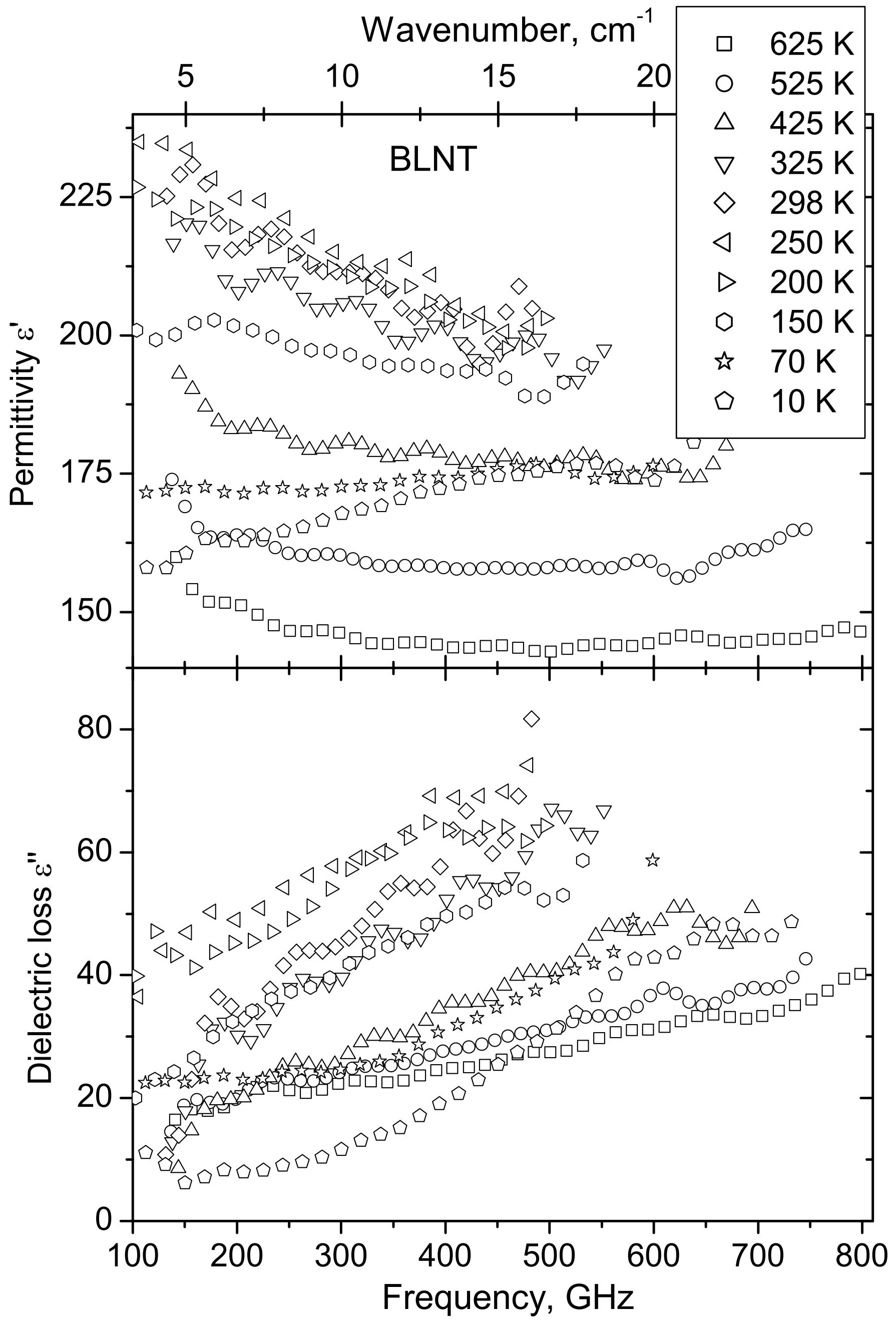

Figure 4

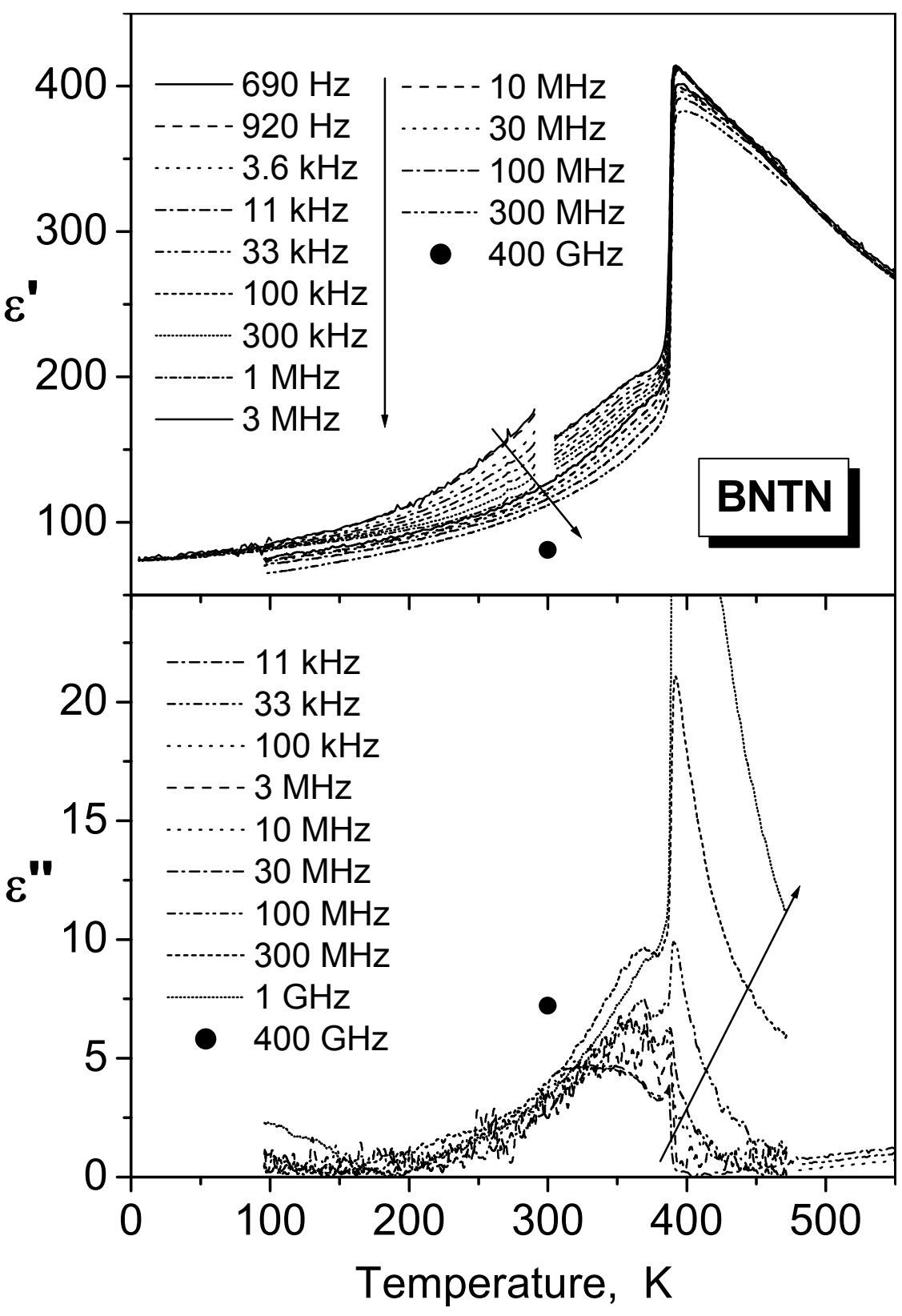

Figure 5

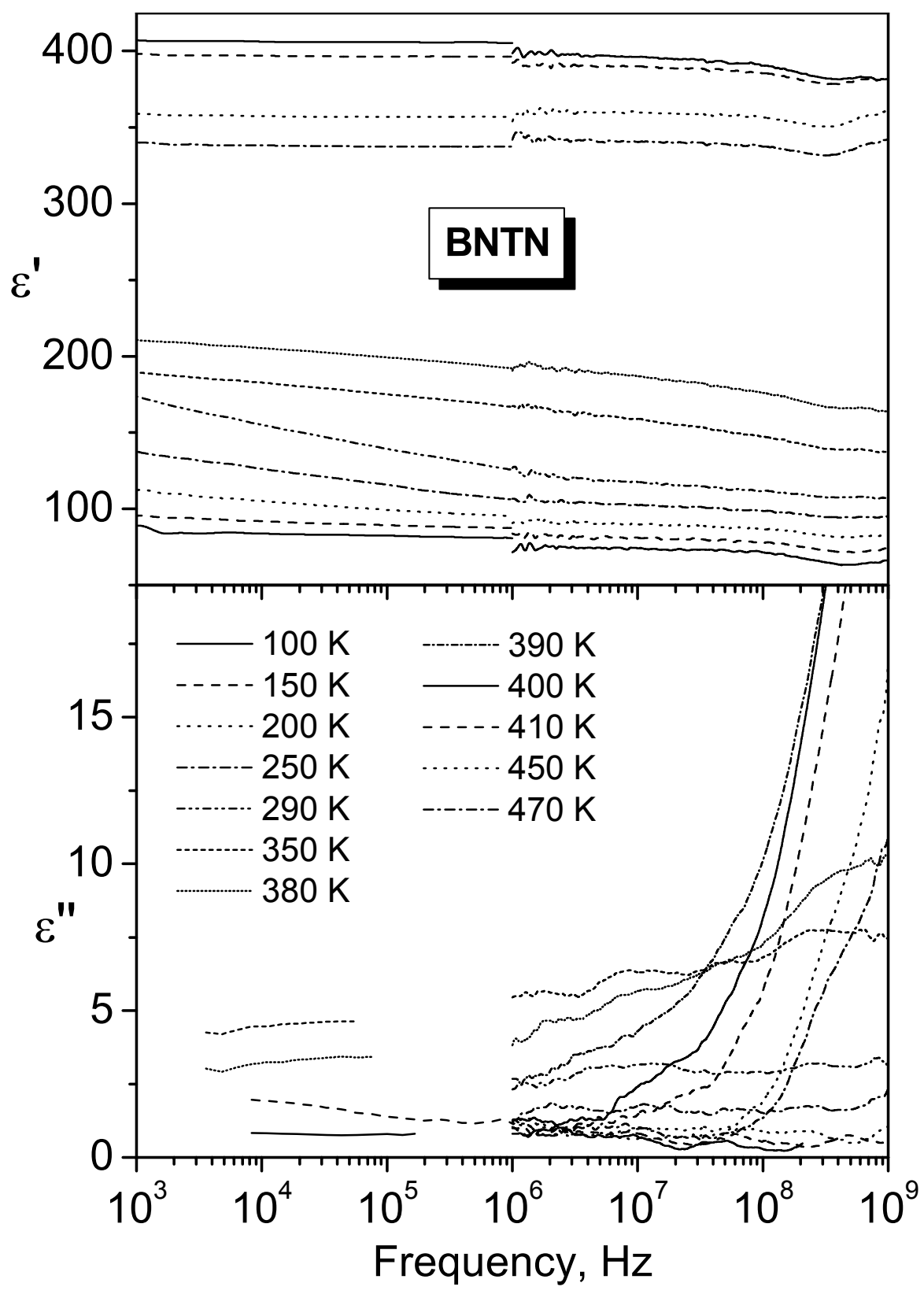

Figure 6